\documentclass[aps,prb,showpacs,twocolumn]{revtex4}
\usepackage{mathrsfs}
\usepackage{amssymb}
\usepackage{amsmath}
\usepackage{bm}
\usepackage{graphics}
\usepackage{natbib}

\begin{document}
\title{Detection of a Majorana-fermion zero mode by a T-shaped quantum-dot structure}

\author{Wei-Jiang Gong$^{1}$}
\author{Shu-Feng Zhang$^{3}$}
\author{Zhi-Chao Li$^{1}$}
\author{Guangyu Yi$^{1}$}
\author{Yi-Song Zheng$^{2}$}
\email[Email address: ] {zys@jlu.edu.cn}
\affiliation{  College of Sciences, Northeastern University, Shenyang 110819, China \\
2. Department of Physics, Jilin University, Changchun 130023, China \\
3. Institute of Physics, Chinese Academy of Sciences, Beijing 10080,
China }
\date{\today}

\begin{abstract}
Electron transport through the T-shaped quantum-dot (QD)
structure is theoretically investigated, by considering a Majorana
zero mode coupled to the terminal QD. It is found that in the
double-QD case, the presence of the Majorana zero mode can
efficiently dissolve the antiresonance point in the conductance
spectrum and induce a conductance peak to appear at the same energy
position whose value is equal to $e^2/2h$. This
antiresonance-resonance change will be suitable to detect the Majorana
bound states. Next in the multi-QD case, we observe that in the
zero-bias limit, the conductances are always the same as the
double-QD result, independent of the parity of the QD number. We
believe that all these results can be helpful for understanding the
properties of Majorana bound states.
\end{abstract}

 \maketitle

\bigskip

\section{Introduction}
Majorana fermions, exotic quasiparticles with non-Abelian
statistics, have attracted a great deal of attention due to both
their fundamental interest and the potential application for the
decoherence-free quantum computation. Different groups have proposed
various ways to realize unpaired Majorana fermions, such as in a
vortex core in a p-wave
superconductor\cite{MBS1,Read,FuL,Sato,Sau,Alicea} or
superfluid.\cite{Kopnin,Sarma} Recently, it has been reported that
Majorana bound states (MBSs) can be realized at the ends of a
one-dimensional p-wave superconductor for which the proposed system
is a semiconductor nanowire with Rashba spin-orbit interaction to
which both a magnetic field and proximity-induced s-wave pairing are
added.\cite{Kitaev,Sau2,Oreg,Wu0} This means that Majorana fermions can
be constructed in solid states, and that its application becomes
more feasible. However, how to detect and verify the existence of
MBSs is a key issue and is rather difficult. Various schemes have
been suggested, including the noise
measurements,\cite{Bolech,Nilsson} the resonant Andreev reflection
by a scanning tunneling miscroscope (STM),\cite{Law} and the $4\pi$
periodic Majorana-Josephson currents.\cite{FuL2}
\par
More recently, some researchers demonstrated that the MBS can be
detected by coupling it laterally to a QD in one closed circuit. The
main reason arises from the quantifiable change of the MBS on the
electron transport through a QD structure. For example, when the QD
is noninteracting and in the resonant-tunneling regime, the MBS
influences the conductance through the QD by inducing the sharp
decrease of the conductance by a factor of $1\over2$, as reported
by D. E. Liu and H. U. Baranger.\cite{Liude} If the QD is in the
Kondo regime, the QD-MBS coupling reduces the unitary-limit value of
the linear conductance by exactly a factor ${3\over4}$.\cite{LeeM}
These results exactly illustrate that the QD structure is a good
candidate for the detection of MBSs. Motivated by these works,
researchers tried to clarify the other underlying transport
properties of the QD structure due to the QD-MBS coupling. Y. Cao
$et$ $al.$ discussed the current and shot noise properties of this
system by tuning the structure parameters.\cite{Lixq} Besides, the
MBS-assisted transport properties have been investigated in the
double-QD structures, and a variety of interesting results have been
observed, such as the crossed Andreev reflection\cite{Zocher} and
nonlocal entanglement.\cite{Hu} These works convinced researchers
that it can be feasible to detect Majorana fermions in the QD
structure. However, the key point is that the experimental results
are difficult to coincide with the value calculated in theory,
because various decoherence factors exist in the experimental
process. This means that it is less convincing to detect the MBSs by
observing the change of resonant tunneling from $e^2\over h$ to $e^2\over 2h$. Therefore, any new
schemes to efficiently detect the MBSs are desirable.
\par
QDs have one important characteristic that some QDs can be coupled
to form the coupled-QD systems. In comparison with the single-QD and
double-QD systems, mutiple QDs present more intricate quantum
transport behaviors, because of the tunable structure parameters and
abundant quantum interference mechanisms. As a typical example, the
antiresonance in electronic transport through a T-shaped multi-QD
structure were extensively studied in the previous
works.\cite{Wangxr,Orellana,Iye,Sato0,Zheng,Torio,Liuy} Such an
effect is tightly related to the parity of QD number. Namely, in the
odd-numbered QD case, resonant tunneling occurs at the low-bias limit. Conversely, for the case of even-numbered QDs, the electronic transport shows the
antiresonance effect which leads to one conductance
zero.\cite{Orellana,Zheng} In view of these results, it is natural
to think that if the MBSs could efficiently modify the transport
properties of the T-shaped QD structure, e.g., the antiresonance effect, such a QD structure will be
a more promising candidate for the detection of MBSs. Motivated by this idea, in the present work we consider a Majorana zero mode to side-couple to the
last QD of the T-shaped QD structure. By calculating the conductance spectrum, we found that the presence of the
Majorana zero mode completely modifies the electron transport
properties of the T-shaped QD structure. The conductance spectra always exhibit the similar conductance peaks
whose values are equal to $e^2\over 2h$ at the zero-bias limit, accompanied by the disappearance of the antiresonance effect. We therefore propose this
structure to be an appropriate candidate to detect the MBSs.

\begin{figure}
\begin{center}\scalebox{0.41}{\includegraphics{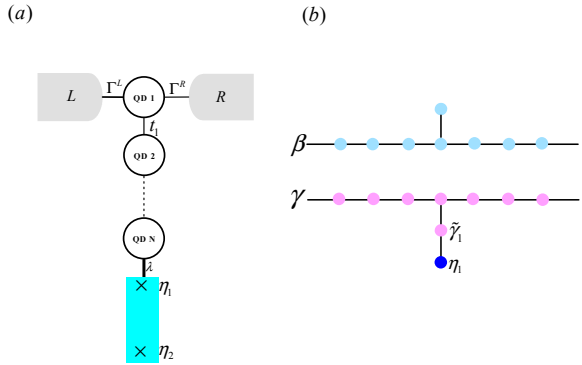}}
\caption{Sketch of a T-shaped QD system with coupled MBSs. The two
MBSs are defined as $\eta_1$ and $\eta_2$, respectively. (b)
Schematic of the T-shaped double-QD structure with coupled MBSs in
the Majorana fermion representation. } \label{Struct}
\end{center}
\end{figure}

\section{model\label{theory}}
The electronic transport structure we propose to detect the MBS is
illustrated in Fig.\ref{Struct}. In such a structure, the last QD of
a noninteracting T-shaped QD system is coupled to one MBS. With the current experimental technique, the T-shaped QD structure
can be readily fabricated. And it is also actually possible to measure its electron transport spectrum. For example, the
antiresonance phenomenon in the electron transport process has been
successfully observed in an recent experimental work.\cite{Japan1} As for the realization of the MBSs, various
schemes have been proposed. For instance, when a semiconductor
nanowire with strong Rashba interaction is subjected to a strong
magnetic field $B$ and adhere to a proximity-induced $s$-wave
superconductivity, a pair of MBSs can form at the end of the
nanowire,\cite{Sau,Oreg} in the case of $V_z
=g\mu_BB/2>\sqrt{\Delta^2+\mu^2}$ ($\Delta$ is the superconducting
order parameter and $\mu$ is the chemical potential of the nanowire).
\par
In Fig.\ref{Struct}, one MBS, defined by $\eta_1$, is assumed to be coupled to QD-$N$. Accordingly, the Hamiltonian of such a structure can be written as
\begin{equation}
H=H_{0}+H_M+H_{MD}.   \label{Hamilt}
\end{equation}
The first term is the Hamiltonian for the T-shaped QD system with
the two connected normal metallic leads, which takes the form as
\begin{eqnarray}
H_{0}&=&\underset{\alpha k }{\sum }\varepsilon _{\alpha k}c_{\alpha
k }^\dag c_{\alpha k }+\sum_{j=1}^{N}\varepsilon _{j}d_{j}^\dag
d_{j} +\sum_{j=1}^{N-1}t_jd_{j}^\dag
d_{j+1}\nonumber\\
&&+\underset{\alpha k }{\sum } V_{\alpha}d_{1}^\dag c_{\alpha k
}+{\mathrm {H.c.}}.\label{2}
\end{eqnarray}
$c_{\alpha k }^\dag$ $( c_{\alpha k })$ is an operator to create
(annihilate) an electron of the continuous state $|k \rangle$ in the
lead-$\alpha$ ($\alpha\in L,R$). $\varepsilon _{\alpha k}$ is the
corresponding single-particle energy. $d^{\dag}_{j}$ ($d_{j}$ ) is
the creation (annihilation) operator of electron in QD-$j$.
$\varepsilon_j$ denotes the electron level in the corresponding QD.
$t_j$ denotes the tunneling between the two neighboring QDs.
$V_\alpha$ is the tunneling element between QD-1 and lead-$\alpha$.
Note that since the QD structure is noninteracting, we in this paper
neglect the spin index. Next, the low-energy effective Hamiltonian
for $H_M$ (i.e., the Majorana fermion) reads
\begin{eqnarray}
H_{M}=i\epsilon_M\eta_1\eta_2.\label{3}
\end{eqnarray}
It describes the paired MBSs generated at the ends of the nanowire
and coupled to each other by an energy $\epsilon_M\sim e^{-l/\xi}$,
with $l$ the wire length and $\xi$ the superconducting coherent
length. The last term in Eq.(\ref{Hamilt}) describes the tunnel
coupling between QD-$N$ and the nearby MBS, which is given by
\begin{eqnarray}
H_{MD}&=&(\lambda d_N-\lambda^* d^\dag_N)\eta_1.\label{TT}
\end{eqnarray}
$\lambda$ is the coupling coefficient between QD-$N$ and the MBS.
\par
By applying a bias voltage $V_b$ between the two leads with
$\mu_L=\varepsilon_F+{eV_b\over2}$ and
$\mu_R=\varepsilon_F-{eV_b\over2}$, we can investigate the electron
transport properties in the presence of Majorana fermion
($\mu_\alpha$ is the chemical potential of lead-$\alpha$, and
$\varepsilon_F$ is the Fermi level in the case of $V_b=0$ which can be assumed to be zero). Note
that in order to realize the robust MBSs, the following condition
must be satisfied: the Zeeman splitting $V_z \gg|V_b|$, $\lambda$,
and $\Gamma$. $\Gamma={1\over2}(\Gamma^L+\Gamma^R)$ is the QD-lead
coupling with $\Gamma^\alpha\equiv 2\pi|V_\alpha|^2\rho$ and $\rho$
the density of states of the leads. One can notice that since the
presence of MBSs, this structure is actually a three-terminal
system. Thus, we have to calculate the current of lead-$L$ and
lead-$R$, respectively, for completely clarifying the electron
transport in this structure. With the help of the nonequilibrium
Green function technique, the current in lead-$\alpha$ is expressed as\cite{Meir}
\begin{equation}
J_\alpha={e\over h}\int d\omega[
T_{ee}^{\alpha\alpha'}(\omega)(f^\alpha_e-f^{\alpha'}_e)+
T_{eh}^{\alpha\alpha}(\omega)(f^\alpha_e-f^{\alpha}_h)].\label{dd}
\end{equation}
In this formula, $f^\alpha_e$ and $f^\alpha_h$ are the Fermi
distributions of the electron and hole in lead-$\alpha$,
respectively. $T_{ee}^{\alpha\alpha'}(\omega)={\rm
Tr}[\Gamma_e^\alpha\textbf{G}^R\Gamma_e^{\alpha'}\textbf{G}^A]$ and
$T_{eh}^{\alpha\alpha}(\omega)={\rm
Tr}[\Gamma_e^\alpha\textbf{G}^R\Gamma_h^{\alpha}\textbf{G}^A]$,
where $\textbf{G}^R$ and $\textbf{G}^A$ are the related and advanced
Green functions. Within the wide-band limit approximation,
$\Gamma^\alpha_e=\Gamma^\alpha_h=\Gamma^\alpha$. Moreover, when the
symmetric-coupling case is considered, i.e., $\Gamma^\alpha=\Gamma$,
the two terms on the right side of Eq.(\ref{dd}) will be equal and
$J_L=-J_R$.
\par
In order to get the
analytical form of the retarded Green function, it is necessary to
switch from the Majorana fermion representation to the completely
equivalent regular fermion one by defining $\eta_1=
(f^\dag+f)/\sqrt{2}$ and $\eta_2=i(f^\dag-f)/\sqrt{2}$ with
$\{f,f^\dag\}=1$. Accordingly, we can write out $H_M$ and $H_D$
respectively as $H_M=\epsilon_M(f^\dag f-{1\over2})$ and
\begin{eqnarray}
H_{MD}= {1\over\sqrt{2}}(\lambda d_N-\lambda^* d^\dag_N)(f^\dag+f).\label{NN}
\end{eqnarray}
\par
Then with the equation of motion method, the matrix form of the retarded Green
function can be written out, i.e.,
\begin{widetext}
\begin{eqnarray}
\textbf{G}^R(\omega)=\left[\begin{array}{ccccccccc}
 g_{1}(z)^{-1} &0&-t_1&0&0&0&0&\cdots &0\\
 0&\tilde{g}_{1}(z)^{-1}&0&t_1&0&0&0&\cdots &0\\
-t^*_1&0&g_{2}(z)^{-1}&0&-t_2&0&0&\cdots &0\\
0&t^*_1&0&\tilde{g}_{2}(z)^{-1}&0&\ddots&0&\cdots&\vdots\\
0&0&-t^*_2& 0&\ddots&0&t_{N-1}&\cdots&\vdots\\
0&0&&\ddots&0&g_{N}(z)^{-1}&0&\frac{\lambda^*}{\sqrt{2}}&\frac{\lambda^*}{\sqrt{2}}\\
\vdots&& & &t^*_{N-1}&0&\tilde{g}_{N}(z)^{-1}&-\frac{\lambda}{\sqrt{2}}&-\frac{\lambda}{\sqrt{2}}\\
0&0&\cdots&0&0&\frac{\lambda}{\sqrt{2}}&-\frac{\lambda^*}{\sqrt{2}}&g_{M}(z)^{-1} &0\\
 0&0&0&0&\cdots&\frac{\lambda}{\sqrt{2}}&-\frac{\lambda^*}{\sqrt{2}}&0&\tilde{g}_{M}(z)^{-1}
\end{array}\right]^{-1}.\ \label{green}
\end{eqnarray}
\end{widetext}
In the above equation,
$g_{j}(z)^{-1}=\omega-\varepsilon_j+i\Gamma\delta_{j1}$ and
$\tilde{g}_{j}(z)^{-1}=\omega+\varepsilon_j+i\Gamma\delta_{j1}$;
$g_{M}(z)^{-1}=\omega-\epsilon_M+i0^+$ and
$\tilde{g}_{M}(z)^{-1}=\omega+\epsilon_M+i0^+$.
Via the above derivation, we can simplify the current formula in this structure as
\begin{equation}
J={e\over h}\int d\omega T(\omega)(f^L_e-f^R_e),
\end{equation}
in which $T(\omega)=-\Gamma{\rm Im} G^R_{11}$.

\section{Numerical results and discussions \label{result2}}
With the formulation developed in the above section, we perform the
numerical calculation to investigate the electron transport
properties of the T-shaped QD structure. In the context, the
symmetric QD-lead coupling is considered, and temperature is fixed
at $k_BT=0$.

\par
\begin{figure}
\begin{center}\scalebox{0.40}{\includegraphics{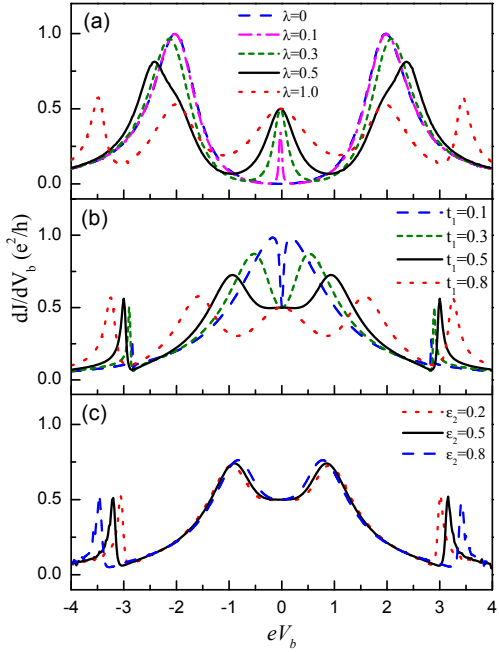}}
\caption{ The conductance spectra of the T-shaped
double-QD structure. The QD-lead coupling is fixed with
$\Gamma={1\over2}$. (a) The conductance as functions of
$eV_b$ with the increase of the coupling between QD-N and $\eta_1$.
The interdot coupling is taken to be $t_1=1.0$. (b) The conductance
as functions changed by the decrease of the interdot coupling.
$\lambda=1.0$. (c) The conductance influenced by the shift of
$\varepsilon_2$ with $t_1={1\over2}$ and $\lambda=1.0$. }
\label{Cond1}
\end{center}
\end{figure}

First of all, we investigate the electron transport properties of
the double-QD configuration with the finite coupling between QD-2
and $\eta_1$. The numerical results are shown in Fig.\ref{Cond1}
where $\varepsilon_j$ is taken to be zero. In Fig.\ref{Cond1}(a), we
find that in the case of $\lambda=0$, the conductance
exhibits two peaks at the points of $eV_b=\pm2.0$, and at the point
of $eV_b=0$ it becomes equal to zero. These two results are easy to
understand. In the case of $\varepsilon_j=0$, the molecular states
of the double QDs are located at the points of $\omega=\pm t_1$.
When $eV_b=\pm 2.0$, the Fermi levels of the leads will coincide
with the energy levels of the molecular states, respectively. On the
other hand, many groups have demonstrated that such a structure
provides two special transmission paths for the quantum
interference. As a result, when the energy of the incident electron
is the same as the energy level of the side-coupled QD, destructive quantum interference will take place, leading to the well-known
Fano antiresonance effect. In the zero-bias limit, only the
zero-energy electron takes part in the quantum transport, so the
conductance zero comes into being.

\par
Next, when the coupling between QD-2 and $\eta_1$ is incorporated,
we can clearly find that the conductance peaks are first suppressed
and then split. What is interesting is that in the presence of
nonzero $\lambda$, the conductance at the zero-bias
point shows a peak. By a further observation, we know that the
conductance value at the energy zero point is exactly equal to
$e^2/2h$. With the enhancement of such a coupling, this conductance
peak is widened, leaving its peak height unchanged. For explaining
this result, we should first solve the value of the conductance peak
mathematically. Based on the expression of $G^R_{11}$ in
Eq.(\ref{G11}), we get the analytical form of $G^R_{11}$ in the
finite-$\lambda$ case, i.e.,
\begin{widetext}
\begin{equation}
G^R_{11}=1/[\omega-\varepsilon_1+i\Gamma-\frac{|t_1|^2\Delta(\omega)-|t_1\lambda|^2\omega(\omega+\varepsilon_1+i\Gamma)}
{(\omega-\varepsilon_2)\Delta(\omega)+|t_1\lambda|^2\omega
-2|\lambda|^2\omega^2(\omega+\varepsilon_1+i\Gamma)}],\label{G11}
\end{equation}
\end{widetext}
where
$\Delta(\omega)=[(\omega+\varepsilon_1+i\Gamma)(\omega+\varepsilon_2)-|t_1|^2]
(\omega^2-\varepsilon_M^2)$. Such a result shows that the nonzero
$\lambda$ indeed complicates the selfenergy of $G^R_{11}$, hence to
modify its properties. It is known that in the case of
$V_b\rightarrow0$, the electron transport is in the linear regime
where $J={\cal G}\cdot V_b$. Here ${\cal G}$ is the so-called linear
conductance defined by ${\cal G}_={e^2\over
h}T(\omega)|_{\omega=0}$. Surely, in such a case, the characteristic
of $G^R_{11}$ in the region of $\omega\rightarrow0$ plays a dominant
role in contributing to the linear conductance. We can readily find
that in the case of $\omega\rightarrow0$, $G^R_{11}$ can be
simplified, i.e.,
\begin{equation}
G^R_{11}\approx{1\over\omega+2i\Gamma}.\label{simple}
\end{equation}
Consequently, the conductance is equal to $e^2\over2h$
in the zero-bias limit.
\par
\begin{figure}
\begin{center}\scalebox{0.40}{\includegraphics{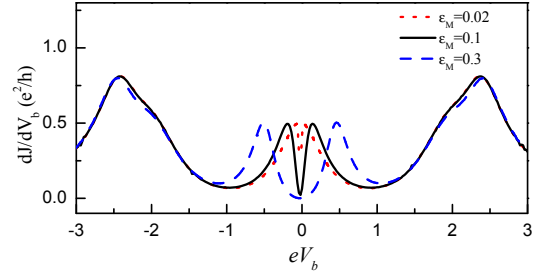}}
\caption{The conductance spectra of the T-shaped double-QD structure
in the case of nonzero coupling between $\eta_1$ and $\eta_2$.}
\label{Finite}
\end{center}
\end{figure}
In order to further analyze the results shown in Fig.\ref{Cond1}(a), we
should clarify the underlying physics mechanism in such a structure.
For this purpose, we rewrite the Hamiltonian in Eq.(\ref{Hamilt}) in
the Majorana representation. To be specific, the two leads should be
first rewritten into two semi-infinite tight-binding fermionic
chains, i.e., $\sum_k\varepsilon _{Lk}c_{Lk }^\dag c_{Lk
}=\sum_{j=-\infty}^{-1}\tau (c_{j}^\dag c_{j-1 }+h.c.)$ and
$\sum_k\varepsilon _{Rk}c_{Rk }^\dag c_{Rk }=\sum_{j=1}^{\infty}\tau
(c_{j}^\dag c_{j+1}+h.c.)$ ( $\varepsilon_{\alpha k}$ and $\tau$ are
confined by the relation of $\varepsilon_{\alpha k}=2\tau\cos k$).
Suppose $d_1=c_0$ ($d^\dag_1=c_0^\dag$), i.e., the two leads with
their connected QD-1 just becomes a one-dimensional chain. Next, by
defining $\beta_j=(c^\dag_j+c_j)/\sqrt{2}$ and
$\gamma_j=i(c^\dag_j-c_j)/\sqrt{2}$, the one-dimensional chain
reduces to two decoupled Majorana chains. By the same token, the
side-coupled QD can be transformed into a MBS by defining
$\tilde{\beta}_1=(d^\dag_2+d_2)/\sqrt{2}$ and
$\tilde{\gamma}_1=i(d^\dag_2-d_2)/\sqrt{2}$. As a consequence, one
can readily find that the T-shaped double-QD structure can exactly
be divided into two isolated T-shaped Majorana chains, as shown in
Fig.\ref{Struct}(b). The difference between these two chains is that
there are two MBSs coupled to each other serially in the lower
branch, whereas in the upper branch only one MBS is presented. For
each branch, the Majorana fermion transport can be evaluated by
means of the nonequilibrium Green function technique. Since the
calculation is simple, we would not like to present the derivation
precess. According to the calculation results, the T-shaped Majorana
chain exhibits the same transport properties as the regular
fermionic one. Namely, when the number of the side-coupled MBSs is
odd, the transport spectra show up as an antiresonance point at the
point of $\omega=0$; instead, the transport will occur resonantly if
the MBS number is even. Therefore, in the T-shaped double-QD
structure with the side-coupled MBSs, the transport is only
contributed by the lower branch. And then, the value of the
conductance is equal to $e^2\over2h$ in the zero-bias
limit.
\par
Fig.\ref{Cond1}(b)-(c) show the influences of changing $t_1$ and
$\varepsilon_2$ on the conductance, respectively. In
Fig.\ref{Cond1}(b), we see that with the decrease of $t_1$, the
conductance peaks in the vicinities of $eV_b=\pm1.5$ enhance and
shift to the zero-bias direction. However, the conductance value at
the zero-bias point is robust with ${\cal G}\equiv {e^2\over2h}$.
Thereby, at such a point the original conductance peak vanishes and
a conductance valley forms. In addition, it can be seen that during
the process of decreasing $t_1$, the conductance peaks around the
points of $eV_b=\pm3.0$ disappears. These results can be understood
as follows. When $t_1$ deceases, QD-2 tends to decouple from QD-1.
In such a case, the strong coupling between QD-2 and $\eta_1$ will
construct a new MBS which couples to QD-1 weakly. Just due to this
reason, we can find that the result of $t_1=0.1$ is consistent with
that of the small $\lambda$ in Ref.\onlinecite{Liude}.
Alternatively, in Fig.\ref{Cond1}(c), it shows that the shift of
$\varepsilon_2$ contributes little to the change of the electron
transport. This is completely opposite to the results in the absence
of MBSs. We can analyze this result with the help of
Eq.(\ref{simple}). We see that in the region of $|\omega|\rightarrow
0$, the terms related to $\varepsilon_2$ are ignored. This exactly
means the trivial role of $\varepsilon_2$. Based on this result, we
readily know that in the presence of MBSs, the fluctuation of QD
levels can not influence the electron transport, which is helpful
for the relevant experiment.
\par
If the MBS wire is not long enough, the two MBSs will be coupled to
each other. In Fig.\ref{Finite} we present the conductance spectra
in the case of nonzero coupling between the two MBSs. It can be
found that different from the results of $\epsilon_M=0$, the nonzero
$\epsilon_M$ induces the appearance of the conductance dip in the
zero-bias limit. When $\epsilon_M=0.02$, the conductance dip is
relatively weak, and the conductance spectrum is consistent with
that in the case of $\epsilon_M=0$ in principle. Next, with the
increase of $\epsilon_M$, the conductance dip becomes apparent.
Especially in the case of $\epsilon_M=0.3$, it exactly becomes an
antiresonance with the wide antiresonance valley. This indicates
that in the case of $\epsilon_M\neq0$, the conductance spectrum will
exhibit an antiresonance point at the zero-bias limit, similar to
the zero-MBS result. However, it should be pointed out that
regardless of the splitting of the conductance peak at the zero-bias
case, the height of the two new conductance peaks near the point of
$V_b=0$ is still close to $e^2\over2h$. Therefore, even not in the
zero mode, the effect of the QD-MBS coupling on the conductance is
distinct.
\begin{figure}
\begin{center}\scalebox{0.41}{\includegraphics{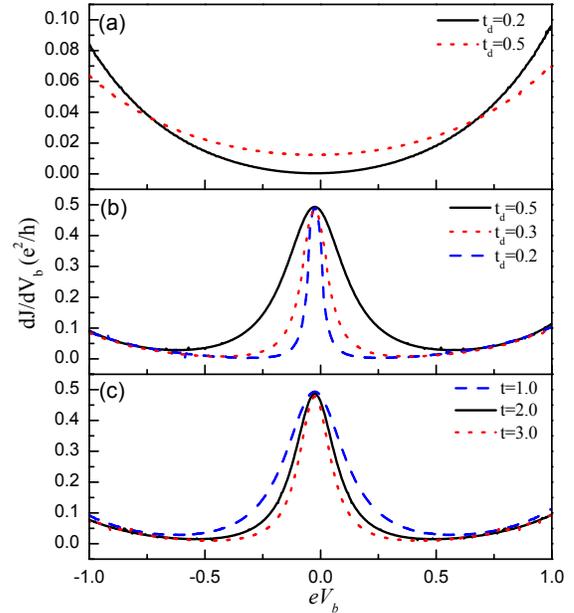}}
\caption{The influence of the Majorana zero mode on the electron transport in the double-QD case when the Majorana zero mode is mimicked by a
semi-infinite chain. The structure parameters are taken as follows: (a) $t=1.0$ and $\Delta=0$; (b) $t=1.0$ and $\Delta=0.3$; (c) $t_d=0.5$ and $\Delta=0.3$. } \label{Real}
\end{center}
\end{figure}
\par
In order to describe the robustness of the MBS signature in the real
physical system, we next calculate the electron transport by writing
the MBS into a one-dimensional semi-infinite topological
superconductor.\cite{Wu1} For simplicity, we write $H_M$ as a
semi-infinite p-wave superconducting chain, i.e.,
$H_M=-\mu\sum_{j}c^\dag_{j}c_{j}+{1\over2}\sum_j[tc^\dag_{j}c_{j+1}+\Delta
e^{i\phi} c^\dag_{j}c^\dag_{j+1} + h.c.]$. Meanwhile, $H_{MD}$ has
its new expression: $H_{MD}=t_d(d^\dag_{2}c_{1}+h.c.)$. By
iteratively solving the end states of the semi-infinite chain, the
MBS-assisted electron transport can be evaluated, and the influence
of the structure parameters of $H_M$ can then be clarified.
Fig.\ref{Real} shows the numerical results with $\mu=0$. In
Fig.\ref{Real}(a), we see that in the case of $\Delta=0$, the
conductance spectra are still characterized by the apparent valleys,
despite the disappearance of the antiresonance. The reason is that
in such a case, the superconductor just becomes a normal electron
reservoir and introduces the inelastic scattering for electron
transmission, hence to weaken the antiresonance effect. But in the
case of $t_d=0.2$, the coupling between QD-2 and the chain is
relatively weak, so that the conductance minimum is almost equal to
zero. On the other hand, in Fig.\ref{Real}(b) when $\Delta=0.3$, one
conductance peak with its value equal to $e^2\over2h$ appears in the
conductance spectra at the zero-bias limit. The decrease of $t_d$
can only narrow the conductance peak but can not suppress its
height. Similar results can be found in the process of increasing
$t$, as shown in Fig.\ref{Real}(c). What is notable in
Fig.\ref{Real}(c) is that when $t$ increases from $2.0$ to $3.0$,
the conductance peak narrows more weakly compared with that of
increasing $t$ from $1.0$ to $2.0$. Based on these results, it can be
found that $\Delta$ and $t_d$ are the two key factors to adjust the
MBS-assisted electron transport. At the same time, these calculations exactly
verify our results in the previous paragraphs.

\par
\begin{figure}
\begin{center}\scalebox{0.40}{\includegraphics{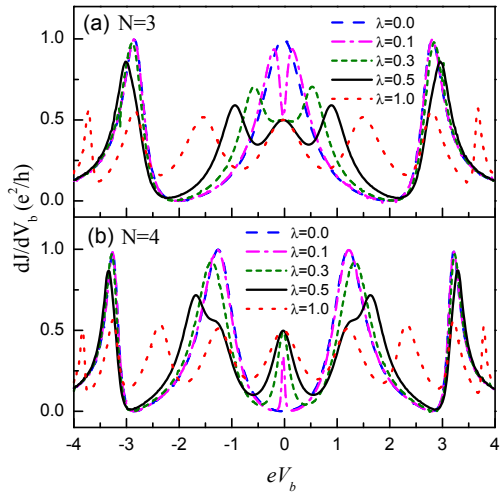}}
\caption{The conductance spectra of the T-shaped multi-QD structure.
In (a) N=3, and N=4 in (b). The relevant parameters are the same as
those in Fig.\ref{Cond1}.} \label{Fanon}
\end{center}
\end{figure}
Motivated by the results of the double-QD structure, we next
investigate the multi-QD case. According to the previous works, the
antiresonance is tightly related to the QD number in the T-shaped
multi-QD structure. Concretely, when the QD number is even,
antiresonance always appear at the zero-bias point; The resonant
tunneling will be observed at such a point
otherwise.\cite{Zheng,Liuy} In Fig.\ref{Fanon} we take the cases of
$N=3$ and $N=4$ to compare the electron transport properties
modified by the MBS in the T-shaped multi-QD structure. From
Fig.\ref{Fanon}(a), we readily find that in the case of $N=3$, the
conductance is equal to $e^2\over h$ around the point
of $eV_b=0$ when $\lambda=0$. When $\lambda=0.1$, despite the weak
coupling between QD-3 and $\eta_1$, the conductance
gradually decreases to $e^2\over2h$ at the zero-bias point.
Consequently, the conductance exhibits a valley around
the zero-bias point. With the increase of $\lambda$, such a valley
becomes widened. When $\lambda$ further increases to $\lambda=0.5$,
the conductance magnitude is suppressed apparently, leading to the
formation of the conductance peak at the zero-bias point. As for the
results in Fig.\ref{Fanon}(b) where $N=4$, we see that they are similar to those
in the double-QD case. The only difference is the increase of the
conductance peaks. These results can be understood by following our
analysis about the double-QD case. In the Majorana fermion
representation, the T-shaped QD structure transforms into two
isolated branches, and the side-coupled MBSs in the two branches
just differ by one. Thus when the transport in one branch is
resonant, the antiresonant transport certainly happens in the other.
Therefore, in the low-bias limit, the conductance is certainly equal
to $e^2\over2h$, independent of the size of the side-coupled QD
chain.
\section{summary\label{summary}}
In summary, we have introduced a Majorana zero mode to couple to the
last QD of the T-shaped QD structure and then investigated the
electron transport in it. After numerical calculation, we have found
that the existence of the Majorana zero mode completely modifies the
electron transport properties of the QD structure. For a typical
structure of double QDs, the coupling between the Majorana zero mode
and the side-coupled QD efficiently dissolves the antiresonance
point in the conductance spectrum and induces a conductance peak to
appear at the same energy position whose value is equal to $e^2\over
2h$. We believe that such an antiresonance-resonance transformation
will more feasible to detect the MBSs, in comparison with the change
of from $e^2\over h$ to $e^2\over 2h$ in the single-QD structure.
Next, the influences of the MBSs on the electron transport in the
multi-QD structure have been discussed. It showed that the
conductance spectra always exhibit the similar conductance peaks
whose values are always equal to $e^2\over 2h$ in the zero-bias
limit, independent of the change of QD number. By transforming the
QD system into the Majorana fermion representation, all the results
have been well clarified. Based on all the obtained results, we
propose that this structure can be a promising candidate for the
detection of the MBSs.
\section*{Acknowledgments}
W. J. Gong thanks B. H. Wu for his helpful discussions. This work
was financially supported by the Fundamental Research Funds for the
Central Universities (Grant No. N110405010), the Natural Science
Foundation of Liaoning province of China (Grants No. 2013020030 and
201202085), and the Liaoning BaiQianWan Talents Program (Grant No.
2012921078).

\clearpage

\bigskip


\begin{thebibliography}{99}

\bibitem{MBS1} G. Moore and N. Read, Nucl. Phys. B \textbf{360}, 362 (1991).

\bibitem{Read} N. Read and D. Green, Phys. Rev. B \textbf{61}, 10267 (2000).

\bibitem{FuL} L. Fu and C. L. Kane, Phys. Rev. Lett. \textbf{100}, 096407 (2008).

\bibitem{Sato} M. Sato, Y. Takahashi, and S. Fujimoto, Phys. Rev. Lett. \textbf{103},
020401 (2009).

\bibitem{Sau} J. D. Sau, R. M. Lutchyn, S. Tewari, and S. Das Sarma, Phys. Rev.
Lett. \textbf{104}, 040502 (2010).

\bibitem{Alicea} J. Alicea, Phys. Rev. B \textbf{81}, 125318 (2010).


\bibitem{Kopnin} N. B. Kopnin and M. M. Salomaa, Phys. Rev. B \textbf{44}, 9667 (1991).

\bibitem{Sarma} S. Tewari, S. Das Sarma, C. Nayak, C. Zhang, and P. Zoller, Phys.
Rev. Lett. \textbf{98}, 010506 (2007).
\bibitem{Kitaev} A. Y. Kitaev, Phys. Usp. \textbf{44}, 131 (2001).

\bibitem{Sau2} R. M. Lutchyn, J. D. Sau, and S. Das Sarma, Phys. Rev. Lett. \textbf{105},
077001 (2010).

\bibitem{Oreg} Y. Oreg, G. Refael, and F. von Oppen, Phys. Rev. Lett. \textbf{105},
177002 (2010).
\bibitem{Wu0} B. H. Wu and J. C. Cao, Phys. Rev. B \textbf{85}, 085415 (2012).

\bibitem{Bolech} C. J. Bolech and E. Demler, Phys. Rev. Lett. \textbf{98}, 237002 (2007).

\bibitem{Nilsson} J. Nilsson, A. R. Akhmerov, and C. W. J. Beenakker, Phys. Rev.
Lett. \textbf{101}, 120403 (2008).

\bibitem{Law} K. T. Law, P. A. Lee, and T. K. Ng, Phys. Rev. Lett. \textbf{103}, 237001
(2009).

\bibitem{FuL2} L. Fu and C. L. Kane, Phys. Rev. B \textbf{79}, 161408 (2009).

\bibitem{Liude} D. E. Liu and H. U. Baranger, Phys. Rev. B \textbf{84}, 201308(R)
(2011).
\bibitem{LeeM} M. Lee, J. S. Lim, and R. L\'{o}pez, Phys. Rev. B \textbf{87},
241402(R)(2013).

\bibitem{Lixq} Y. Cao, P. Wang, G. Xiong, M. Gong, and X. Q. Li, Phys. Rev. B
\textbf{86}, 115311 (2012).

\bibitem{Zocher} B. Zocher and B. Rosenow, Phys. Rev. Lett. \textbf{111},
036802 (2013).

\bibitem{Hu} Z. Wang, X. Y. Hu, Q. F. Liang, and X. Hu, Phys. Rev. B \textbf{87}, 214513
(2013).


\bibitem{Wangxr} X. R. Wang, Yupeng Wang, and Z. Z. Sun, Phys. Rev. B \textbf{65}, 193402
(2002).

\bibitem{Orellana} P. A. Orellana, F. Dom\'{\i}nguez-Adame, I. G\'{o}mez, and M. L.
Ladr\'{o}n de Guevara, Phys. Rev. B \textbf{67}, 085321 (2003).

\bibitem{Iye} K. Kobayashi, H. Aikawa, A. Sano, S. Katsumoto, and Y. Iye, Phys.
Rev. B \textbf{70}, 035319 (2004).

\bibitem{Sato0} M. Sato, H. Aikawa, K. Kobayashi, S. Katsumoto, and Y. Iye, Phys.
Rev. Lett. \textbf{95}, 066801 (2005).

\bibitem{Zheng} Y. Zheng, T. L\"{u}, C. Zhang, and W. Su, Physica E (Amsterdam)
\textbf{24}, 290 (2004).

\bibitem{Torio} M. E. Torio, K. Hallberg, S. Flach, A. E. Miroshnichenko, and M.
Titov, Eur. Phys. J. B \textbf{37}, 399 (2004).

\bibitem{Liuy} Y. Liu, Y. Zheng, W. Gong, T. L\"{u}, Phys. Lett. A \textbf{360}, 154 (2006).


\bibitem{Japan1} S. Sasaki, H. Tamura, T. Akazaki, and T. Fujisawa
Phys. Rev. Lett. \textbf{103}, 266806 (2009); M. Sato, H. Aikawa, K. Kobayashi, S. Katsumoto, and Y. Iye,
Phys. Rev. Lett. \textbf{95}, 066801 (2005).

\bibitem{Meir} Y. Meir and N. S. Wingreen, Phys. Rev. Lett. \textbf{68}, 2512
(1992); W. Gong, Y. Zheng, Y. Liu, and T. L\"{u}, Phys. Rev. B
\textbf{73}, 245329 (2006).


\bibitem{Wu1} A. C. Potter and P. A. Lee, Phys. Rev. Lett. \textbf{105}, 227003
(2010).
\end{thebibliography}
\end{document}